\documentclass[a4paper,twoside]{article}

\usepackage{array}
\usepackage{hyperref}
\usepackage{amsmath}
\usepackage{graphicx}
\usepackage{subcaption}
\usepackage{calc}
\usepackage{amssymb}
\usepackage{amstext}
\usepackage{amsmath}
\usepackage{amsthm}
\usepackage{multicol}
\usepackage{pslatex}
\usepackage{apalike}
\usepackage{color}
\usepackage{xcolor}
\usepackage{hyperref}

\usepackage[bottom]{footmisc}
\usepackage{SCITEPRESS}     
\hypersetup{
    colorlinks,
    linkcolor={black!50!black},
    citecolor={black!50!black},
    urlcolor={black!80!black}
}

\definecolor{linkblue}{rgb}{0,0,0.7} 
\begin{document}

\title{Network Analysis of the Egyptian Reddit Community}

\author{
\authorname{Samy Shaawat \sup{1}\orcidAuthor{0009-0002-2679-6695}, 
            Adham Hammad\sup{1}\orcidAuthor{0009-0005-4587-7310} , 
            Karim Farhat \sup{1}\orcidAuthor{0009-0001-4777-4201},
            Mina Thabet  \sup{1}\orcidAuthor{0009-0005-7368-7207} and 
            Walid Gomaa  \sup{1,2}\orcidAuthor{0000-0002-8518-8908}}
\affiliation{\sup{1}Faculty of Engineering,
                    Department of Computer Science Engineering, 
                    Egypt-Japan University of Science and Technology, 
                    Alexandria, 
                    Egypt
                    }               
\affiliation{\sup{2}Faculty of Engineering, 
                    Alexandria University, 
                    Alexandria, 
                    Egypt
                    }              
\email{\{samy.shaawat, adham.khalid, karim.farhat, mina.thabet, walid.gomaa\}@ejust.edu.eg}
}


\abstract{This paper presents a network analysis of the Reddit community focused on Egypt. We collected and constructed a comprehensive dataset consisting of 23,185 users and 105 Egyptian subreddits. Through network analysis criteria such as degree analysis, degree distribution analysis, and clustering coefficient analysis, we explored the structural properties, connectivity patterns, and local clustering within the Egyptian Reddit network. The findings provide insights into the community dynamics, influential users, and information flow within the network. Our study contributes to a better understanding of online communities in the context of Egypt and sheds light on the relationships and interactions within the Egyptian Reddit community. By leveraging network analysis techniques, we uncover the importance of individual nodes, the distribution of node degrees, and the formation of tightly knit groups.This study contributes significantly to the understanding of online communities specific to Egypt, shedding light on relationships and interactions within the Egyptian Reddit community. }

\onecolumn \maketitle \normalsize \setcounter{footnote}{0} \vfill

\section{\uppercase{Introduction}}
\label{sec:introduction}

\par 
Social media platforms have become an integral part of our lives, enabling us to connect, share, and learn from each other. Among these platforms, Reddit stands out as one of the most popular and influential online communities in the world.\cite{DigitalTrends22}

\par 
Reddit is a social bookmarking website that allows users to submit, rate, and comment on various types of content, such as news, images, videos, and text posts. Users can join and create subreddits, which are specialized forums dedicated to specific topics or interests. According to Semrush, Reddit had more than 430 million monthly active users as of October 2021, making it the fourth most visited site in the U.S. and the sixth most visited worldwide.\cite{Tomsguide21}

\par 
Reddit is a rich source of data and insights for researchers who want to study online communities, social network analysis, and user behavior. However, most of the existing studies on Reddit have focused on the global or English-speaking subreddits, while neglecting the regional or non-English subreddits that represent diverse and vibrant communities around the world. This oversight hampers our understanding of the unique dynamics and contributions of these regional communities.

Our research addresses this gap by focusing on the Egyptian Reddit community, a dynamic online community centered around topics related to Egypt. This community comprises several subreddits covering various aspects of Egyptian culture, politics, society, and entertainment.\cite{RedditSecrets21}

\par
The research problem at the heart of our study is to investigate the structure and dynamics of the Egyptian Reddit community, which is largely unexplored in the existing literature. Specifically, we aim to understand how this community functions, the patterns of user interaction within it, and its role in shaping discussions related to Egypt.

\par
Our study aims to address the research problem of the largely unexplored Egyptian Reddit community by investigating its structure, dynamics, and contributions. Specifically, we seek to analyze the network structure of this community, identifying key nodes, influencers, and clusters of interest. Additionally, we aim to examine user behavior patterns within the Egyptian Reddit community, including content sharing, engagement, and the dissemination of information. Furthermore, our research endeavors to understand the cultural significance of Egyptian Reddit subreddits by investigating how they contribute to discussions on various aspects of Egyptian culture, politics, society, and entertainment, both within the community and in the broader online discourse.~\cite{STOLTENBERG2019120}

\par 
The paper is organized as follows. Section~\ref{sec:introduction} is an introduction. Section ~\ref{sec:Literature Review} gives a literature review about on social network analysis, Reddit analysis, and network analysis in online communities. 
Section~\ref{sec:Methodology} presents the core of our work given the overall methodology, an overview of the network, showing the connections between users in the Egyptian Reddit community through visual 
representation. We discuss the network's construction and provide information about the number of users and shared subreddits. In Section~\ref{sec:Results and Discussion} we give our empirical work and the
corresponding analyses. Section~\ref{sec:conclusion} concludes the paper with pointers to future work.


\section{\uppercase{Literature Review}}
\label{sec:Literature Review}
\par
Social media networks like Reddit, Facebook, and Twitter contain a wealth of data that can provide insights into human behavior and social connections.~\cite{BuiltIn21}

\subsection*{Community Identification on Reddit}
\label{subsec:Community Identification on Reddit}
\vspace{-8pt}
Smith et al. analyzed the network structure of Reddit to identify different types of user communities. Their purpose was to understand the diverse connections and communities on Reddit. This research provides a general overview of how people interact and form communities on the platform, and it is easily understandable for general readers.~\cite{Smith19}

\subsection*{User Engagement in Reddit Communities}
\label{subsec:User Engagement in Reddit Communities}
\vspace{-8pt}
Xu et al. studied the degree distributions of Reddit communities to understand their connectivity patterns. They wanted to explore the range of user engagement levels within Reddit communities. By doing so, they provide insights into how active users are in different communities on Reddit.~\cite{Xu19}

\subsection*{Temporal Dynamics of Twitter Networks}
\label{subsec:Temporal Dynamics of Twitter Networks}
\vspace{-8pt}
Johnson examined how networks change over time on Twitter. The purpose was to understand the dynamics of social media networks and how some communities persist while others are more temporary. This research showcases the constant evolution of social media networks in a way that is easy to grasp.~\cite{Johnson20}

\subsection*{Connectivity Patterns in Twitter Hashtags}
\label{subsec:Connectivity Patterns in Twitter Hashtags}
\vspace{-8pt}

Park et al. analyzed local and global connectivity patterns in the Twitter hashtag network. They aimed to provide an intuitive understanding of how hashtags are connected on Twitter, demonstrating the layered connectivity patterns within the platform.~\cite{Park20}

\subsection*{Age-Related Differences in Facebook Connections}
\label{subsec:Age-Related Differences in Facebook Connections}
\vspace{-8pt}
Cho and Lee studied clustering coefficients across different Facebook networks to understand how human connections differ across age groups on social media. Their purpose was to examine the differences in how people of different ages connect on Facebook. This research offers insights into age-related differences in social media usage.~\cite{Cho21}

\subsection*{Political Polarization on Brexit-Related Facebook Pages}
\label{subsec:Political Polarization on Brexit-Related Facebook Pages}
\vspace{-8pt}
Williams and Housley analyzed Facebook pages related to Brexit to understand the level of polarization in communities. They wanted to examine how political polarization manifests in social networks. This work provides a straightforward look at the ways political polarization is reflected in online communities.~\cite{Williams21}

\subsection*{Global Events and Reddit Community Connections}
\label{subsec:Global Events and Reddit Community Connections}
\vspace{-8pt}
Ahmed et al. analyzed the network structure of Reddit discussions about current events in Egypt. Their purpose was to understand how world events shape social connections on platforms like Reddit. This research offers an accessible view of how global events influence online communities.~\cite{Ahmed22}

\subsection*{Idea Spread in Reddit Discussions}
\label{subsec:Idea Spread in Reddit Discussions}
\vspace{-8pt}
Xu and Ke used natural language processing techniques to extract topics from Reddit comments and constructed networks connecting comments discussing the same topics. They aimed to understand how ideas spread and become popular within Reddit. This research gives insights into the dynamics of idea sharing within the platform.~\cite{Xu22}

\subsection*{Cross-Platform Social Media Engagement}
\label{subsec:Cross-Platform Social Media Engagement}
\vspace{-8pt}
Lee et al. took a cross-platform view and constructed social media networks that connected the same users across multiple platforms. Their purpose was to understand how individuals engage across different social media websites. This work provides an intuitive understanding of how people connect and interact on various platforms.~\cite{Lee23}  \\

These previous studies provide insights into human behavior and social connections on social media platforms. They explore various aspects such as community formation, user engagement, network dynamics, connectivity patterns, age group differences, political polarization, and the impact of world events.

\section{\uppercase{Methodology}}
\label{sec:Methodology}
\subsection{Dataset}
\label{subsubsec:Dataset}

\par 
In this section, we will explain the process of preparing and constructing a dataset focused on Egyptian subreddits. Our goal was to create a comprehensive dataset for analyzing discussions and content related to Egypt, utilizing the open API provided by Reddit and the Python PRAW library.

\subsubsection{Dataset Collection and Preprocessing}

\par 
The initial step in dataset preparation involved collecting data from Reddit, specifically targeting Egyptian subreddits.  
\vspace{2pt}

\noindent {\bf Identification of Egyptian Subreddits:}  Using the open Reddit API and the Python PRAW library, we conducted a search for subreddits related to Egypt. Our search query looked for subreddits where the term 'Egypt' (in English) appeared in either the name or the content. This search resulted in the retrieval of 105 relevant subreddits, which we saved for further processing.

\vspace{2pt}
\noindent {\bf Extraction of Usernames from Subreddits:} Using the authenticated Reddit API, we proceeded to extract active usernames from the collected Egyptian subreddits. We iterated through each subreddit and retrieved the submissions, including both posts and comments. From these submissions, we extracted the usernames of the authors. We included only those usernames that were associated with at least one submission within the subreddit, ensuring that we captured usernames of active participants. We then removed duplicate usernames, resulting in a final list of unique usernames associated with the Egyptian subreddits. We saved this list in a text file named after the corresponding subreddit. The size of the dataset is shown in table ~\ref{tab:Dataset Size}, and the details of the dataset are discussed in the next section.

\begin{table}[h]
\caption{Dataset Size}\label{tab:Dataset Size} \centering
\begin{tabular}{|c|c|}
    \hline
        Number of rows  & 23,184 \\
    \hline
        Number of Columns & 24\\
    \hline
\end{tabular}
\end{table}

\subsubsection{Dataset Construction}

\par 
To construct the dataset, we generated text files, with each file representing a subreddit and containing the active usernames associated with that subreddit.
Next, we iterated over all the text files, extracting usernames and linking them to the subreddits where they were active. Each username was assigned its own column in the dataset, and in the respective column, we marked the subreddits where the user was active.

\par 
The dataset consists of 23,185 rows as shown in the pervious table ~\ref{tab:Dataset Size}, each representing a unique username. The dataset comprises 24 columns, with the first column dedicated to usernames, while the remaining 23 columns correspond to the subreddits that each username is subscribed to. The reason for the 23 columns, rather than the total number of subreddits (105) in the dataset, is that no username is subscribed to all 105 subreddits. The maximum number of subreddits to which a username is subscribed is 23. As a result, the dataset is structured to include only the columns to represent the subreddit subscriptions of each username.

\par 
Once the dataset was created, we conducted a thorough quality check to ensure data integrity. We examined the dataset, searching for null values or duplicates. If any were found, we promptly removed them from the dataset. This process was crucial in ensuring the reliability and accuracy of the dataset. \footnote{\href{https://github.com/SamyShaawat/Algorithm-Project-ALG-S2023-07.git}{\color{linkblue}{Click to see our Dataset and Network Implementation}}\label{Dataset Repo}}

\subsection{Network Overview}
\label{subsubsec:Network Overview}

\par
\begin{figure}[!h]
  \centering
   {\includegraphics[width = 5cm]{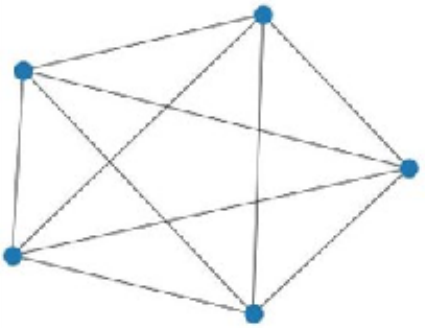}}
  \caption{Connected component number 10. It represents the subreddit named ``EgyptianHistoryMemes'' and consists of 5 members: IacobusCaesar, Joseph-Memestar, RoroS4321, AnticRetard, and Memetaro-Kujo.}
  \label{fig:Component10}
 \end{figure}

 
 The network constructed from the dataset provides a visual representation of the connections between users based on their shared subreddit interests. The nodes represent individual users, while the undirected edges depict the presence of common subreddits between pairs of users. There are no weights assigned to the edges, and whether there are many or just one subreddit in common, they are handled in the same way. This network structure enables the analysis of relationships and information flow within the Egyptian Reddit community fig.~\ref{fig:Component10} shows a sample of a network constructed using the collected dataset.

\subsubsection{Network Construction}

\par
\begin{figure}[!h]
  \centering
   {\includegraphics[width = 6cm]{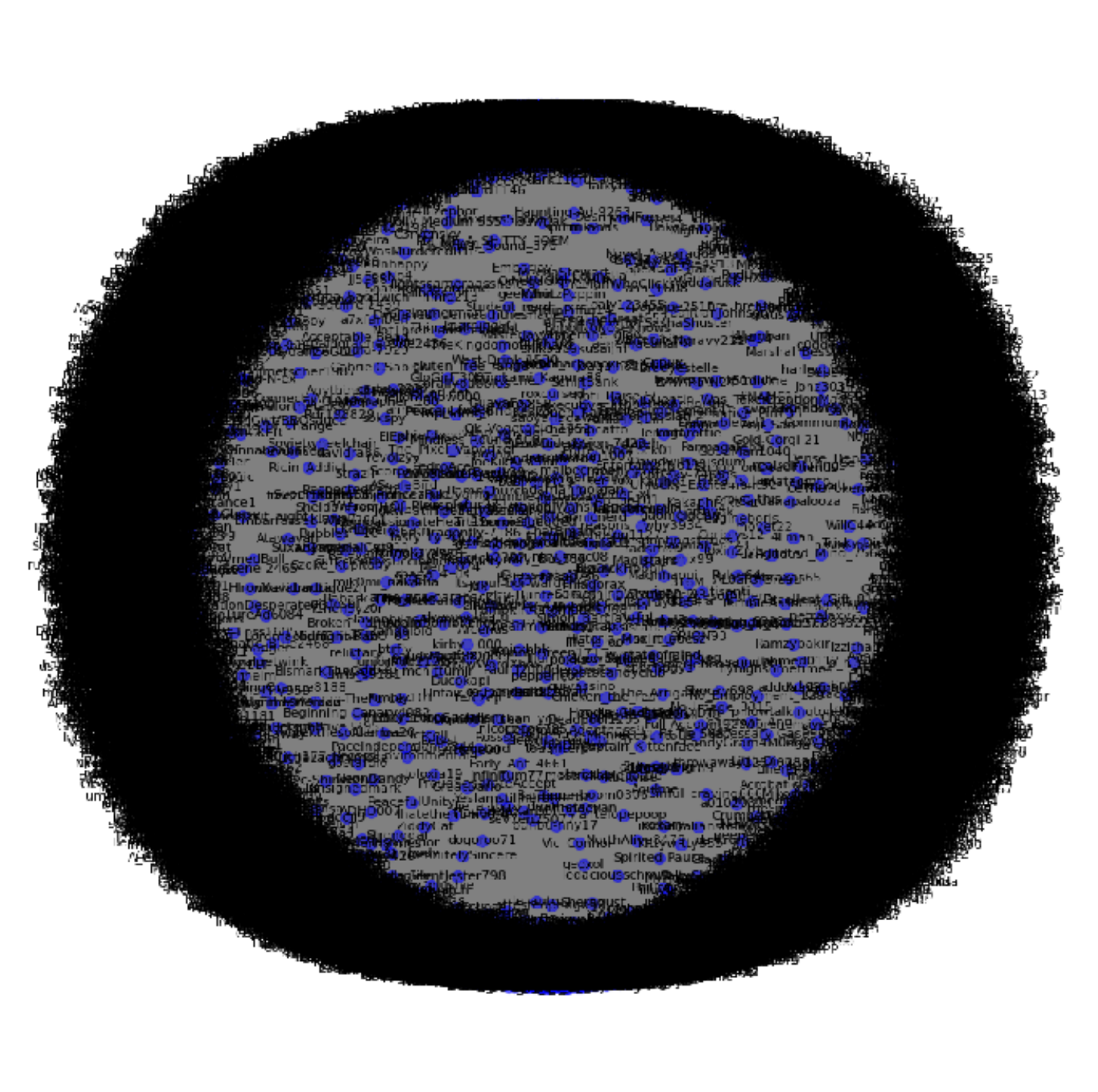}}
  \caption{Fully Generated Network (with labels)}
  \label{fig:LabelsNetwork}
\end{figure}

\begin{figure}[!h]
  \centering
   {\includegraphics[width = 6cm]{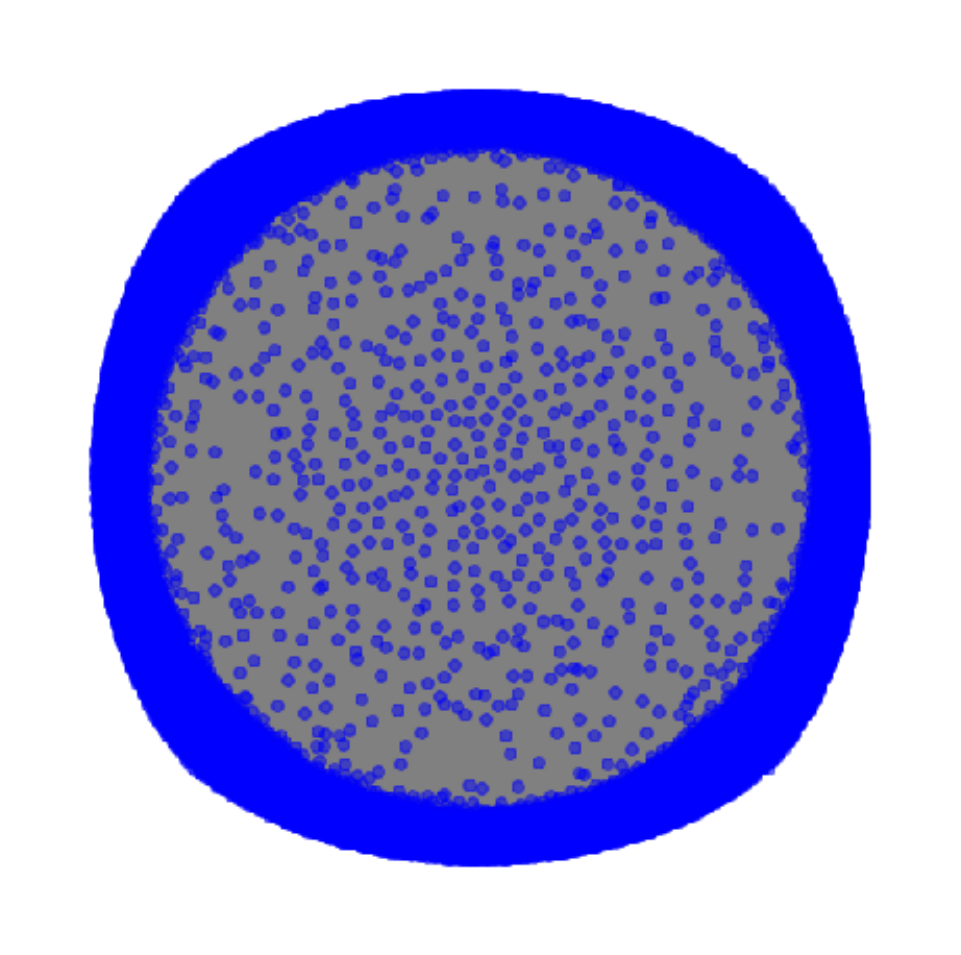}}
  \caption{Fully Generated Network (without labels)}
  \label{fig:NOLabelsNetwork}
\end{figure}

 
The network is constructed using python modules: Pandas, NetworkX, and Matplotlib. 
The Network successfully constructed as shown in fig.~\ref{fig:LabelsNetwork} and ~\ref{fig:NOLabelsNetwork} to analyze the structure and dynamics of the Egyptian Reddit community. This network, comprised of 23,185 users and 105 subreddits~\footref{Dataset Repo}, forms the basis for further exploration and investigation into users interactions, influential nodes, and community dynamics~\cite{PowellHopkins20}.

\subsubsection{Network Metrics}

\par 
The following are some metrics we used to analyze the network.
\begin{enumerate} 
    \item Node Count: The dataset comprises 23,185 unique users (nodes) who follow Egyptian subreddits. These users form the nodes of the network.
    \item Edge Count: The edges in the network indicate the presence of shared subreddits between pairs of users. The total number of edges in the network
    represents the level of interconnectedness, overlap in subreddit interests among users. There are a total of \noindent {\bf 6,877,773} edges.
\end{enumerate}

\begin{table}[h]
\caption{Number of Nodes and Edges.}\label{tab:Network Metrics} \centering
\begin{tabular}{|c|c|}
    \hline
        Network Metrics & Value \\
    \hline
        Number of Nodes & 23,185\\
    \hline
        Number of Edges & 6,877,773\\
    \hline 
    
\end{tabular}
\end{table}

\subsection{Network Analysis Criteria}

\par 
Network analysis is a powerful methodology for studying and understanding complex systems represented as graphs or networks. It involves analyzing the structure, relationships, and dynamics of nodes and edges within the network to gain insights into the underlying system's behavior and characteristics. Network analysis encompasses various techniques and measures that provide valuable information about connectivity, centrality, community structure, and other properties of the network~\cite{PowellHopkins15}.

\subsubsection{Degree Analysis}

\par 
Degree analysis, also known as degree centrality, is a network analysis measure that focuses on the number of connections or edges that a node (in this case, Egyptian Reddit users) has in the network. It helps identify highly connected users who follow a large number of subreddits, indicating their active engagement and broad range of interests within the Reddit community. Degree analysis provides insights into the overall network structure, popular subreddits, and influential users, revealing patterns of information sharing and community interactions \cite{Degree}.

\par

\subsubsection{Degree Distribution Analysis}

\par 
Degree distribution analysis in the network analysis of Reddit for Egyptians examines the distribution of node degrees, which represent the number of connections or edges that users have in the network. By analyzing the frequency or probability distribution of these degrees, we can understand the prevalence and distribution of user engagement and participation within the Egyptian Reddit network. Skewness and kurtosis measures provide valuable insights into the connectivity patterns, centralization, and structural characteristics of the network, revealing the concentration of highly connected users and the overall distribution of degrees.To calculate the average degree for undirected network, let $N$ be the number of nodes, and $L$ be the number of edges; then the expected node degree can be calculate as shown in Eq.~\ref{eq1} ~\cite{Degree}.

\par

\begin{equation}\label{eq1}
    <K> = \frac{\sum deg(i)}{N}=\frac{2L}{N}
\end{equation}

\subsubsection{Clustering Coefficients}

\par 
Clustering coefficient analysis provides insights into the local connectivity and clustering tendencies within the Egyptian Reddit community. It helps identify tightly interconnected clusters or communities within the network, highlighting users who actively engage and form connections with fellow Egyptian users. The clustering coefficient analysis allows us to understand the local connectivity, community structures, and interaction patterns within the Egyptian Reddit community ~\cite{CLEMENTE201826}.

\subsubsection{Network Type}

\par 
Network type analysis categorizes the network based on its structural properties. This analysis helps us understand the fundamental characteristics of the network, such as connectivity patterns and overall structure. Common network types include random networks, small-world networks, clustered networks, and sparse networks. By identifying the network type, we gain insights into the connectivity patterns, community structures, and the behavior within the Egyptian Reddit community ~\cite{article}.

\subsubsection{Centrality Analysis}

\par 

Centrality analysis in the network analysis of Reddit for Egyptians helps identify important nodes based on measures like Degree centrality, Closeness centrality, Betweenness centrality, and Eigenvector centrality. Degree centrality assesses the number of connections, Closeness centrality measures accessibility, Betweenness centrality identifies bridge nodes, and Eigenvector centrality considers connections with influential nodes. Centrality analysis reveals key nodes that play significant roles in the structure and information flow of the Egyptian Reddit network ~\cite{inbook}.

\subsubsection{Static Community Discovery}

\par 
Community discovery involves partitioning the network into groups of nodes called communities. These communities consist of nodes that have stronger connections or similarities within the same community compared to nodes in different communities. Community detection algorithms are applied to identify these meaningful communities, providing insights into the modular structure and functional units within the Egyptian Reddit network. The quality of the detected communities is evaluated using metrics such as modularity and conductance. Visualizations aid in interpreting the communities, revealing relationships and interactions between nodes. Further analysis involves studying the characteristics and functions of nodes within each community, contributing to a comprehensive understanding of the network's structure and dynamics ~\cite{ZHU2020127}.

\subsubsection{Dynamic Community Discovery}

\par 
Dynamic community discovery network analysis of Reddit for Egyptians involves identifying and tracking communities in evolving graphs that represent the interactions between users over time. This analysis captures the changing structure and temporal evolution of communities within the Egyptian Reddit network. Various approaches, such as the Label Propagation Algorithm, graph partitioning techniques, the Louvain Method, and the Infomap algorithm, can be employed to address the challenges of dynamic community discovery. These approaches help uncover how communities form, evolve, and interact within the evolving Egyptian Reddit network~\cite{thompson2017static}.

\subsubsection{Connected Components Analysis}

\par 
Connected components analysis in the network analysis of Reddit for Egyptians identifies distinct clusters or subgraphs within the larger network. It partitions nodes into subsets called connected components, where nodes within a component are mutually reachable. This analysis helps reveal the underlying structure, connectivity patterns, and isolated regions within the Egyptian Reddit network. Determining component sizes and visualizing the components provide insights into cluster distribution and the main connected part of the graph, contributing to a better understanding of connectivity and identifying isolated regions or clusters~\cite{HUANG2009173}.

\subsubsection{Density Analysis}

\par
Density analysis in the network analysis of Reddit for Egyptians quantifies the level of connectivity or sparsity within the Egyptian Reddit network. It measures the ratio of the number of edges present in the graph to the maximum possible edges as shown in Eq.~\ref{eq2}. Higher density values indicate a more densely connected network, while lower density values suggest a more sparse or fragmented network. Density analysis provides insights into the overall connectivity and structure of the Egyptian Reddit network, highlighting the level of engagement and information flow within the community~\cite{GOSWAMI201816}.
\begin{equation}\label{eq2}
    DensityRatio = \frac{\#Edges}{\#Nodes * (\#Nodes - 1) / 2} 
\end{equation}

\subsubsection{Path Analysis}

\par 
Path analysis in the network analysis of Reddit for Egyptians involves studying the routes between nodes to understand connectivity, reachability, and flow within the network. It includes finding the shortest path, analyzing reachability, studying path length distribution, examining flow and traffic patterns, and uncovering connectivity patterns like hubs and bridges. Path analysis provides valuable insights into the structure and dynamics of the Egyptian Reddit network, aiding in understanding information flow and communication within the community~\cite{thompson2017static}. Let $N$ be the number of nodes in the network, then the expected path distance can be calculated as shown in Eq.~\ref{eq3}

\begin{equation}\label{eq3}
    <D> = \frac{\sum dist(i,j)}{{N \choose 2}}
\end{equation}

\section{\uppercase{Results and Discussion}}
\label{sec:Results and Discussion}
\subsection{Degree Analysis}
\label{subsubsec:Degree Analysis}

\par 
The degree of a node in a graph is the number of edges connected to that node. In our study, we analyzed the degrees of nodes in the graph and made the observation prrovided in table.~\ref{tab:Degree Analysis}.

\begin{table}[h]
\caption{Degree general statistics}\label{tab:Degree Analysis} \centering
\begin{tabular}{|c|c|}
    \hline
        Observation & Value \\
    \hline
        Minimum degree	 & 0\\
    \hline
        Maximum degree & 13,019 \\
    \hline
        Average degree & 593.6\\
    \hline
        Total number of degrees & 137,555,46\\
    \hline
\end{tabular}
\end{table}

\par 
The degree analysis of the network revealed interesting observations. The minimum degree of 0 indicated the presence of isolated nodes, while the maximum degree of 13019 pointed to highly connected nodes or hubs. The average degree of 593.602 reflected the typical connectivity of nodes, and the total number of degrees was 13,755,546, illustrating the overall size and complexity of the network. These findings offer insights into the network's structure and potential real-world applications. To deepen our understanding, we ask: How does the degree distribution compare to other regional Reddit communities, and what role do these highly connected nodes play in information flow and community cohesion within this network?

\subsection{Degree Distribution Analysis}
\label{subsubsec:Degree Distribution Analysis}

\par 
The degree distribution analysis examines the variation and patterns of node connectivity in a network. It provides insights into influential nodes, information flow, and network structure. This analysis helps uncover the distribution characteristics of degrees in the network.

\begin{figure}[h]
  \centering
   {\includegraphics[width = 7cm]{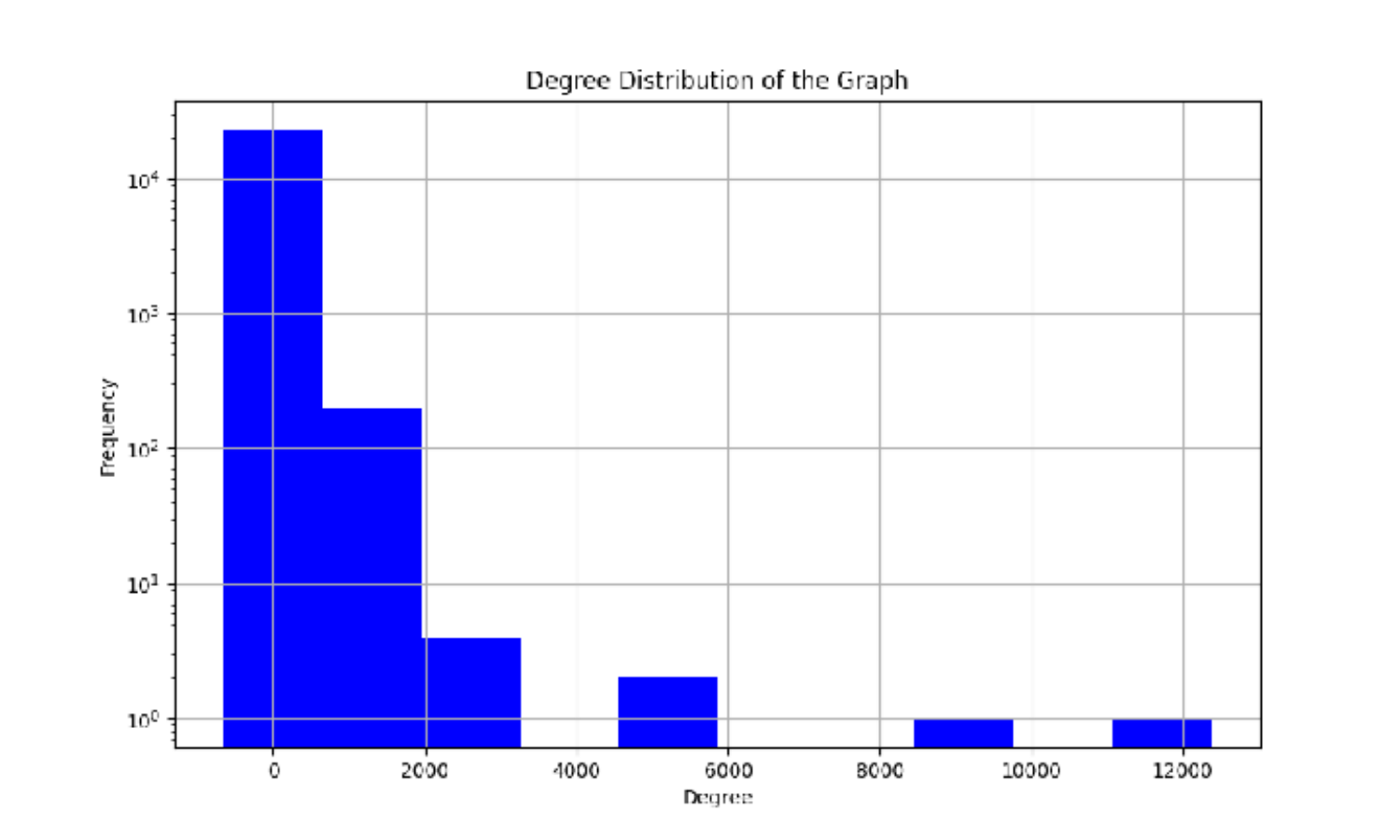}}
  \caption{Degree distribution of the Egyptian reddit network.}
  \label{fig:DegreeDistribution}
\end{figure}

\begin{table}[h]
\caption{Degree distribution general statistics}\label{tab:Degree Distribution Analysis} \centering
\begin{tabular}{|c|c|}
    \hline
        Observation & Value \\
    \hline
        Range of Degree Distribution & 0 to 13,019\\
    \hline
        Average Degree & 593.6 \\
    \hline
        Standard Deviation & 257.98\\
    \hline
        Skewness & Positive\\
    \hline
        Kurtosis & Leptokurtic\\
     \hline
    
\end{tabular}
\end{table}

\par 
The degree distribution analysis reveals important characteristics of the network's connectivity patterns as provided in table.~\ref{tab:Degree Distribution Analysis}. The observed range of degrees from 0 to 13,019 indicates substantial variation in the number of edges per 
node. The average degree of 593.602 with a standard deviation of 257.982 signifies the typical number of connections and the degree of variation. The positively skewed distribution suggests that most nodes have low degrees, while a few nodes exhibit high degrees, highlighting an uneven distribution of edges. And this is as well indicated in the histogram shown in fig.~\ref{fig:DegreeDistribution}. Additionally, the leptokurtic distribution shows a sharp peak and heavy tails, indicating a concentration of nodes around the average with a presence of nodes with significantly higher degrees. This analysis not only help us comprehend network properties but also lay the groundwork for addressing the following research questions: How does this degree distribution compare to other regional Reddit communities, and what role do highly connected nodes play in influencing network dynamics, including information dissemination efficiency?

\subsection{Clustering Coefficients}
\label{subsubsec:Clustering Coefficients}

\par 
The clustering coefficient ranges from 0 to 1, where 0 means that none of the node’s neighbors are connected to each other and 1 means that all the node’s neighbors are connected to each other.
The average clustering coefficient ranges from 0 to 1, where 0 means that there is no clustering at all and 1 means that there is perfect clustering.
The average clustering coefficient of the graph was calculated to be $0.976$.
Average clustering and average degree were used to determine Network Type.

\begin{figure}[h]
  \centering
   {\includegraphics[width = 7cm]{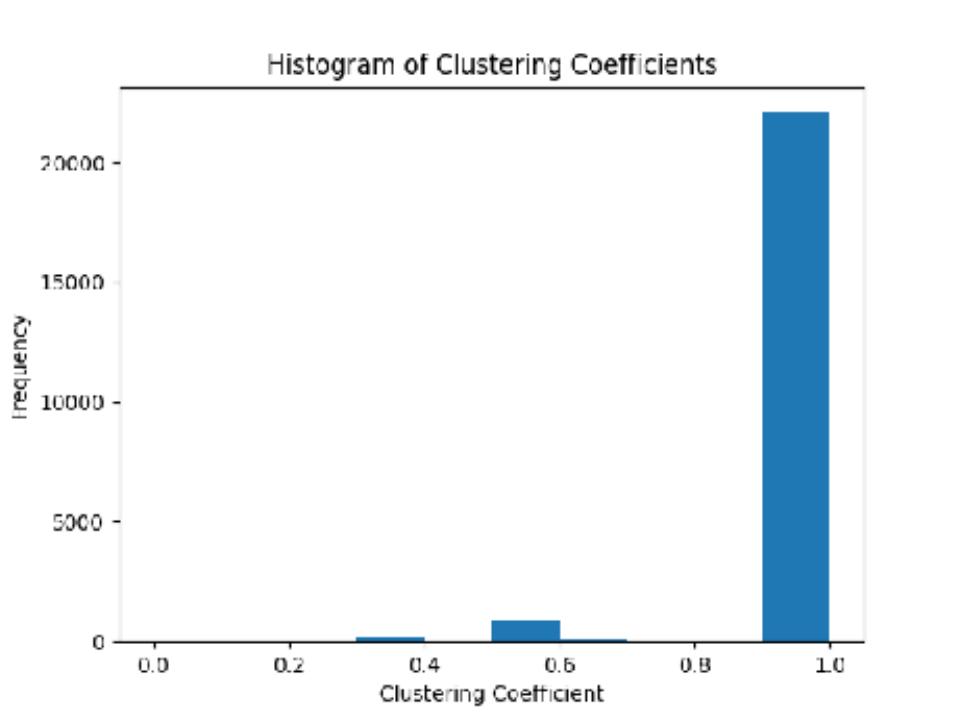}}
  \caption{Histogram of clustering coefficients for the Egyptian reddit network.}
  \label{fig:ClusteringCoefficients}
\end{figure} 

\par 
Our observation is that the graph exhibits a remarkably high level of clustering as shown in fig.~\ref{fig:ClusteringCoefficients}, with the majority of nodes forming triangles with their neighbors, indicating strong local connectivity and the presence of tightly-knit communities. The presence of numerous cliques or communities further underscores the pronounced clustering in the graph. Additionally, the graph's connectivity patterns suggest a departure from randomness or sparsity, as there are abundant common friends among the nodes, implying a dense network structure with interconnected nodes. This observation highlights the cohesive nature of the graph, with distinct communities and cohesive clusters contributing to its overall structure and connectivity. 
The high level of clustering observed in the network of the Egyptian Reddit community could be attributed to factors such as the shared cultural and linguistic background of the users, as well as potential offline interactions and shared experiences. The clustering effect may be intensified by the predominance of Arabic subreddits, fostering stronger connections among users with similar interests and cultural affiliations.

\subsection{Network Type}
\label{subsubsec:Network Type}

As mentioned before the degree distribution and clustering coefficient can provide insights into the network type, our observations are shown in table~\ref{tab5:Network Type}. following are the measures used to infer the network type:

\subsubsection{Degree Distribution}  

\par 
Start by calculating the degree of each node, which represents the number of edges connected to it. Then, examine the degree distribution, which shows the frequency distribution of node degrees. If the degree distribution follows power-law distribution which implies a "rich-get-richer" phenomenon, where a small number of nodes acquire a disproportionately large number of connections while the majority of nodes have relatively fewer connections, it indicates a scale-free network. On the other hand, if the degree distribution is more uniform or bell-shaped, it suggests a random network where connections formed randomly, no specific pattern or structure,relatively uniform distribution of connections and no dominant hubs or regular network where there is a structured pattern of connections, equal number of connections for each node and often forms clusters or neighborhoods, respectively.

\subsubsection{Clustering Coefficient:}  
\par 
Calculate the clustering coefficient for each node, which measures the tendency of nodes to cluster together. Compute the average clustering coefficient for the entire graph by averaging the clustering coefficients of all nodes. If the average clustering coefficient is much higher than expected in a random graph with the same size and degree distribution, it indicates highly clustered or small-world network in which tightly connected clusters and short distances between nodes exist and it enables efficient communication. If it's close to the expected value of a random graph, it suggests a random network where connections between nodes are formed randomly without any specific pattern and a close-to-zero average clustering coefficient suggests a sparse or low clustering network which has few connections between nodes and limited clustering and it has a low density of connections, long average distances between nodes and limited inter-connectivity.

\begin{table}[h]
\caption{Network type of the Egyptian reddit network.}\label{tab5:Network Type} \centering
\begin{tabular}{|c|c|}
    \hline
        Observation & Value \\
    \hline
        Network clustering & High\\
    \hline
        Network connectivity & High\\
    \hline
        Network Type & Small World Network.\\
    \hline    
    
\end{tabular}
\end{table}

\subsection{Centrality Analysis}
\label{subsubsec:Centrality Analysis}

\par 
In this section, we present a centrality analysis of the Egyptian reddit network, focusing on a user named \noindent {\bf Wil}, who exhibits remarkable characteristics in terms of his \noindent Degree Centrality, Betweenness Centrality, Closeness centrality, and Eigenvector centrality. Wil is a member of our dataset, specifically identified in row 19206. Our observations indicate that Wil is highly influential and well connected within the network. 
\par
Wil's centrality within the network can be assessed through various measures as provided in table.~\ref{tab5:Centrality Analysis}, providing insights into his active engagement and influential role. Degree centrality, quantified as 0.561842, indicates that Wil follows 23 out of the 105 subreddits in the network, highlighting his active participation and interest in a significant portion of the network's content. Moving on to betweenness centrality, Wil's score of 0.269488 suggests his crucial position as a bridge connecting different parts of the network, facilitating communication and the flow of information between nodes. This implies that he plays a vital role in maintaining the network's connectivity and enabling efficient information dissemination. Wil's high closeness centrality, measured at 0.692209, demonstrates his proximity to other nodes in the network, allowing information to quickly spread through the network via him and enabling efficient dissemination of information. Moreover, Wil's eigenvector centrality score of 0.036501 highlights his connections to other influential nodes in the network, enhancing his potential for influence within the network.

\begin{table}[h]
\caption{Centrality metrics of the user named \bf Wil.}\label{tab5:Centrality Analysis} \centering
\begin{tabular}{|c|c|}
    \hline
        Centrality Metric & Value \\
    \hline
        Degree centrality & 0.561842\\
    \hline
        Betweenness centrality & 0.269488\\
    \hline
        Closeness centrality & 0.692209\\
    \hline
        Eigenvector centrality	 & 0.036501\\
    \hline 
    
\end{tabular}
\end{table}

\subsection{Static Community Discovery}
\label{subsubsec:Community Discovery}

\par 
Static Community discovery was performed using rigorous methodology. Community detection algorithms, specifically Louvain or Girvan-Newman, were employed to identify distinct communities within the network. The quality of 
the communities was evaluated using modularity, while community sizes and overlap were analyzed. Where available, the detected communities were validated against groundtruth. For evolving networks, temporal dynamics were considered, and the results were interpreted within the network's context. The analysis was refined iteratively based on network characteristics.

\par
\begin{figure}[h]
  \centering
   {\includegraphics[width = 7cm]{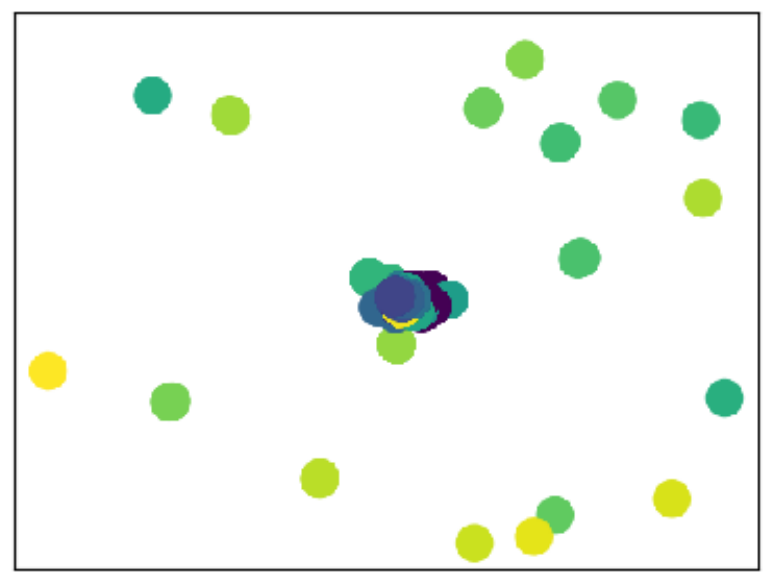}}
  \caption{50 (Static) communities in the Egyptian Reddit network}
  \label{fig:Static Community Discovery}
\end{figure}

 
We observed that the Louvain algorithm, emerges as a widely used community detection algorithm that effectively identifies communities in networks. By iteratively optimizing the modularity measure, which quantifies the quality of network division, the Louvain algorithm demonstrates its ability to partition networks into cohesive communities as shown in fig.~\ref{fig:Static Community Discovery}. This observation underscores the algorithm's significance in the field of network analysis, where it provides a valuable tool for uncovering underlying community structures within complex networks.

\begin{table}[h]
\caption{Static Community Discovery}\label{tab:Static Community Discovery} \centering
\begin{tabular}{|>{\centering\arraybackslash}p{17mm}|>{\centering\arraybackslash}p{13.5mm}|>{\centering\arraybackslash}p{17mm}|>{\centering\arraybackslash}p{13.5mm}|}
    \hline
        Static Community no. & No. of Elements & Static Community no. & No. of Elements\\
    \hline
        0 & 217 & 25 & 559 \\
    \hline
        1 & 758 & 26 & 709 \\
    \hline
        2 & 512 & 27 & 368 \\
    \hline
        3 & 717 & 28 & 751 \\
    \hline
        4 & 916 & 29 & 509 \\
    \hline
        5 & 800 & 30 & 1 \\
    \hline
        6 & 617 & 31 & 1 \\
    \hline
        7 & 416 & 32 & 24 \\
    \hline
        8 & 785 & 33 & 1 \\
    \hline
        9 & 625 & 34 & 4 \\
    \hline
        10 & 2228 & 35 & 9 \\
    \hline
        11 & 777 & 36 & 1 \\
    \hline
        12 & 1316 & 37 & 1 \\
    \hline
        13 & 638 & 38 & 5 \\
    \hline
        14 & 491 & 39 & 5 \\
    \hline
        15 & 728 & 40 & 2 \\
    \hline
        16 & 1267 & 41 & 88 \\
    \hline
        17 & 399 & 42 & 3 \\
    \hline
        18 & 732 & 43 & 1 \\
    \hline
        19 & 766 & 44 & 5 \\
    \hline
        20 & 809 & 45 & 1 \\
    \hline
        21 & 952 & 46 & 1 \\
    \hline
        22 & 751 & 47 & 1 \\
    \hline
        23 & 746 & 48 & 637 \\
    \hline
        24 & 522 & 49 & 1 \\
    \hline

\end{tabular}
\end{table}

\par
In fig.~\ref{fig:Static Community Discovery} the colored circles represent 50 communities (provided in table.~\ref{tab:Static Community Discovery}) in the Egyptian reddit network and the distance between the circles shows how different the sizes of these communities are. When circles are close together, it means the communities have a similar number of users, mostly consisting of a large number of usernames and if the circles are far apart, it means there is a big difference in the number of users between those communities, typically representing communities with a small number of usernames.

\subsection{Dynamic Community Discovery}
\label{subsubsec:Dynamic Community Discovery}

\par 
Dynamic community discovery was conducted using a systematic approach. Algorithms designed for dynamic networks were utilized, applying them to each time slice with a defined temporal resolution. Community 
stability, persistence, evolution, and changes were analyzed, considering temporal metrics for additional insights.

\par 
We observed that Label Propagation algorithm, stands out as an efficient approach for community detection in graphs, compared to other algorithms, it is computationally more efficient, flexible in handling different graph types, and does not require prior knowledge of the number of communities. However, it may be sensitive to noise and highly connected nodes and may not capture global community structure as effectively. 
This semi-supervised algorithm leverages the network structure and neighbor labels to assign community labels to nodes. By propagating labels iteratively throughout the network, the algorithm reaches a stable state where each node is assigned a label that maximizes agreement with its neighbors. The Label Propagation algorithm's ability to exploit local connectivity patterns and iteratively refine community assignments highlights its effectiveness as a powerful tool for identifying communities in graphs.

\par
\begin{figure}[h]
  \centering
   {\includegraphics[width = 7cm]{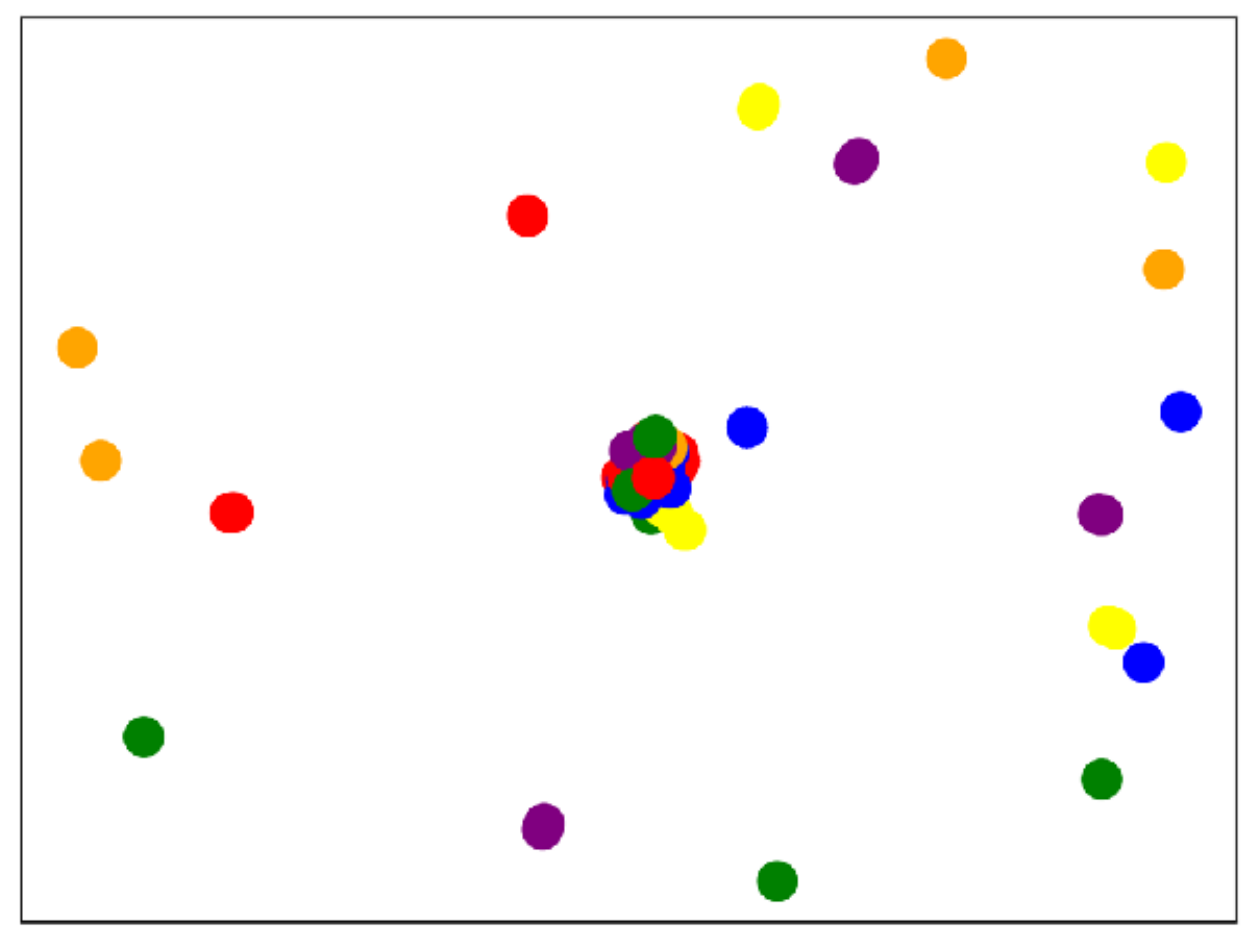}}
  \caption{39 (Dynamic) Communities in Egyptian reddit network}
  \label{fig:Dynamic Community Discovery}
\end{figure}

The results were visualized as shown in fig.~\ref{fig:Dynamic Community Discovery}, the colored circles represent 39 communities (provided in table.~\ref{tab:Dynamic Community Discovery}) in the Egyptian reddit network. The arrangement and distance between the circles show how the communities change over time or in different states. When circles are close together, it means the communities at that time or state have a similar number of users, mostly consisting of a large number of usernames. If the circles are far apart, it means there is a significant difference in the number of users between those communities, typically representing communities with a small number of usernames.

\begin{table}[h]
\caption{Dynamic Community Discovery}\label{tab:Dynamic Community Discovery} 
\centering
\begin{tabular}{|>{\centering\arraybackslash}p{17mm}|>{\centering\arraybackslash}p{13.5mm}|>{\centering\arraybackslash}p{17mm}|>{\centering\arraybackslash}p{13.5mm}|}
    \hline
        Dynamic Community No. & No. of Elements & Dynamic Community No. & No. of Elements\\
    \hline
        0 & 13468 & 21 & 16 \\
    \hline
        1 & 8189 & 22 & 3 \\
    \hline
        2 & 368 & 23 & 1 \\
    \hline
        3 & 3 & 24 & 5 \\
    \hline
        4 & 1 & 25 & 5 \\
    \hline
        5 & 66 & 26 & 3 \\
    \hline
        6 & 28 & 27 & 1 \\
    \hline
        7 & 2 & 28 & 9 \\
    \hline
        8 & 1 & 29 & 1 \\
    \hline
        9 & 24 & 30 & 2 \\
    \hline
        10 & 1 & 31 & 15 \\
    \hline
        11 & 4 & 32 & 2 \\
    \hline
        12 & 9 & 33 & 9 \\
    \hline
        13 & 1 & 34 & 1 \\
    \hline
        14 & 1 & 35 & 153 \\
    \hline
        15 & 5 & 36 & 243 \\
    \hline
        16 & 208 & 37 & 12 \\
    \hline
        17 & 5 & 38 & 1 \\
    \hline
        18 & 2 & 39 & 217 \\
    \hline
        19 & 88 &  & \\
    \hline
        20 & 1 &  &  \\
    \hline
    
\end{tabular}
\end{table}

\vfill
\subsection{Connected Components Analysis}
\label{subsubsec:Connected Components Analysis}

\par 
The graph analysis reveals the presence of 19 connected components, indicating that the graph is divided into distinct groups of nodes with each forming a connected component, a sample of connected component no. 10 is shown in fig.~\ref{fig:Component10}. These components are separate entities and lack direct connections between them. Among these components, Component 1 stands out with the maximum number of elements, reaching 23,042. Notably, the elements within component 1 do not belong to the same subreddit, suggesting a diverse composition. Additionally, the analysis highlights that 11 components consist of a single element, while 8 components comprise more than one element. This observation with components sizes' provided in table.~\ref{tab:Connected Components Analysis} significantly impacts the density ratio, emphasizing the variation in component sizes and the presence of both isolated and interconnected substructures within the graph.

\begin{table}[h]
\caption{Connected Components Analysis}\label{tab:Connected Components Analysis} \centering
\begin{tabular}{|c|c|c|c|}
    \hline
        Component & Size & Component & Size \\
    \hline
        1 & 23042 & 11 & 2\\
    \hline
        2 & 1 & 12 & 88\\
    \hline
        3 & 1 & 13 & 3\\
    \hline
        4 & 1 & 14 & 1\\
    \hline
        5 & 4 & 15 & 1\\
    \hline 
        6 & 9 & 16 & 5\\
    \hline
        7 & 1 & 17 & 1\\
    \hline
        8 & 1 & 18 & 1\\
    \hline
        9 & 5 & 19 & 1\\
    \hline
        10	 & 5 & &\\
    \hline 
\end{tabular}
\end{table}

\vfill
\subsection{Density Analysis}
\label{subsubsec:Density Analysis}

\par 
Density is a measure of how connected a graph is, calculated as the ratio of the number of edges to the number of possible edges in a graph.

\begin{equation}\label{eq4}
    \# Possible Edges = \frac{(\#Nodes) * (\#Nodes - 1)}{2}
\end{equation}
\begin{equation}\label{eq5}
    Density Ratio = \frac{\#Edges}{\#Possible Edges}
\end{equation}

\par 
The analysis of the network reveals that it is relatively sparse with fewer edges compared to a fully connected network. 
This is supported by the higher count of disconnected components compared to connected components. The network consists of 23,173 nodes and 6,877,773 edges, indicating the relationships between these entities. The density ratio, calculated as 0.025617 using Eq.~\ref{eq5}, confirms the sparse nature of the network, with lower values indicating fewer connections. These observations shown in table.~\ref{tab:Density Analysis} are crucial for understanding the network's connectivity patterns and analyzing the data represented by its nodes and edges.

\begin{table}[h]
\caption{Density Analysis Observations}\label{tab:Density Analysis} \centering
\begin{tabular}{|c|c|}
    \hline
        Observation & Value \\
    \hline
        Network sparsity & Relatively sparse \\
    \hline
        Number of Nodes	 & 23,173 \\
    \hline
       Number of Edges & 6,877,773 \\
    \hline    
        Number of possible edges & 268,482,378 \\
    \hline
        Density Ratio & 0.025617 \\
    \hline
    
\end{tabular}
\end{table}

\subsection{Path Analysis}
\label{subsubsec:Path Analysis}

\par 
Many algorithms have been developed for path analysis, but among these techniques the most important and popular one is the shortest path analysis. Shortest path is implemented by using the connected components output file and randomly choose two different usernames from each component. In total there are 19 components, 10 of which each has one element, so they were ignored and we focused on 9 other components which have more than one element. As shown in table.~\ref{tab:Path Analysis}:
component 1, along with components 5, 6, 9, 10, 11, 12, 13, and 15, have more than one username, these components differ from the others, which have only one username. Component 1 is unique because it contains usernames from multiple subreddits, while the other components have usernames within a single subreddit and this distinction is important because it affects the path length between usernames. In component 1, the path length can be equal to 2 because there may be an intermediate node between the source and target usernames, so this means there is an extra step in the path due to the multiple subreddit connections within component 1.

In contrast, the other components have a path length of 1 because the usernames are all in the same subreddit. This direct path does not require any intermediate nodes between the source and target usernames.

\begin{table}[h]
\caption{Shortest path results}\label{tab:Path Analysis} \centering
\resizebox{\textwidth/2}{!}{
\begin{tabular}{|c|c|c|c|}
    \hline
        Component & Source & Target & Shortest Path Length\\
    \hline
        1 & gqn & cLoUt diDDit & 2\\
    \hline
        5 & kendralinnette & odedi1 & 1  \\
    \hline
        6 & happyboy13 & ProgaPanda & 1  \\
    \hline    
        9 & fatrna ezzouhry & AEssam & 1  \\
    \hline
        10 & lacobusCaesar & Memetaro Kujo & 1  \\
    \hline
        11 & lnTheKurry & RealHistoryMashup & 1  \\    
    \hline
        12 & dishonoredgraves & LimeAndTacos & 1  \\
    \hline
        13 & CASCADE 999 & Trainer Opposite & 1  \\
    \hline
        15 & muffled savior & 3pr7 & 1 \\
    \hline

\end{tabular}
}
\end{table}

\section{\uppercase{Conclusions}}
\label{sec:conclusion}

\par 
In this research paper we focused on analyzing the Reddit network for Egyptians. We followed a step-by-step methodology to collect and preprocess the dataset, resulting in a comprehensive dataset with 23,185 unique users and 105 Egyptian subreddits.
The network constructed from the dataset provided a visual representation of the connections between users based on their shared subreddit interests. With 6,877,773 edges, the network showed a significant level of interconnectedness among users.

\par 
Through the application of various network analysis techniques, such as degree analysis, degree distribution analysis, clustering coefficient analysis, and network type analysis, we gained insights into the characteristics of the network. These analyses helped us understand the degrees of nodes, the distribution of degrees, the level of clustering within the network, and the network's overall structural properties.

\par 
Our research contributes significantly to the body of knowledge surrounding the Egyptian Reddit community, enriching our understanding of its dynamics and underlying structures. These findings hold practical relevance for the identification of influential users, the study of information propagation, and the exploration of community frameworks. Researchers and community managers alike can leverage these insights to make more informed decisions and foster more effective engagement within the Egyptian Reddit community.

\par 
However, it is essential to acknowledge the limitations of our study. While our analysis provides a robust foundation for understanding the Egyptian Reddit network, the dynamic nature of online communities means that our findings may not be static over time. Future research endeavors should consider incorporating temporal analysis to capture the evolving nature of the network. Furthermore, the incorporation of sentiment analysis could unveil the emotions and opinions expressed within Egyptian subreddits, adding depth to our understanding.

\par
Looking beyond the confines of this study, similar analyses can be extended to other communities and populations, such as the broader Arab and African communities, shedding light on the evolving landscape of underprivileged countries and regions.

\par
At the end, our research offers valuable insights into the Egyptian Reddit network, illuminating its inner workings and serving as a launchpad for future studies and community engagement efforts. By addressing the unique characteristics of this regional Reddit community, we advance the frontiers of knowledge in online community analysis and offer a roadmap for future investigations in this dynamic field.


\bibliographystyle{apalike}
{\small
\bibliography{References}
}
\end{document}